\documentclass[12pt]{article}

\title{Klein--Gordon Equation for Quark Pairs in Color Superconductor}
\author{B.O.Kerbikov\\ Institute of Theoretical and Experimental
Physics\\ Moscow, Russia}
\date{}
\begin{document}

\maketitle

{\small The wave equation is derived  for quark pairs in color
superconductor in the regime  of low density/strong coupling.}

\vspace{2cm}

 \large

 During the last five  years color superconductivity
became a compelling topic in QCD --  see the review papers
\cite{1}. Grossly speaking, we have a fair understanding of color
superconductivity physics in the high density/weak coupling
regime. In the low density/strong coupling region the situation is
different. Here, the theory faces the well-known difficulties of
the nonperturbative QCD  and use is made of the models, like NJL,
or instanton gas. By low density we mean the quark densities 3--4
times larger than that in the normal nuclear matter. Model
calculations show (see \cite{2} and references in \cite{1}) that
at such densities the  gap equation acquires a nontrivial
solution. This was interpreted \cite{1} as the onset of the
Bardeen--Cooper--Schrieffer (BCS) regime, i.e., the formation of
the condensate of Cooper pairs made of $u$- and $d$-quarks. It is
known, however, that nonzero value of the gap is only a signal of
the presence of the fermion pairs. Depending on the dynamics of
the system, on the fermion density, and on the temperature, such
pairs may be either stable, or fluctuating in time, may form a BCS
condensate, or the dilute Bose gas, or undergo a Bose
condensation. The continuous evolution from the BCS regime to the
regime of the Bose--Einstein condensation (BEC) is called the
BCS--BEC crossover. Such a transition takes place either by
increasing the strength of the interaction, or by decreasing the
carrier density. The fact that the BCS wave function may undergo a
smooth evolution and describe the tightly bound fermion pairs was
first noticed long ago [3--6]. According to \cite{5}, the remark
that ``there exist a transformation that carries the BCS into BE
state'' was originally made by F.J.Dyson in 1957 (i.e., the same
year when the BCS paper \cite{7} was published). The BCS--BEC
crossover for quarks was first discussed in \cite{8}. The general
description of the crossover for quarks will be given elsewhere.

In the present paper we shall derive an equation for the quark
pairs in the low density/stong coupling limit. This equation is
obtained directly from the mean-field gap equation. Our derivation
follows the work by P.~Nozieres and S.~Schmitt-Rink \cite{4}, who
obtained the Schr\"{o}dinger equation starting from the BCS
solution. Quarks in color superconductor are relativistic
particles, and therefore we shall arrive to the Klein--Gordon
equation.

Our starting point is the expression for the thermodynamic
potential for the 2SC superconductor. In the 2SC phase \cite{1}
pairing takes place between $u$- and $d$-quarks, while $s$-quark
is out of the game until the density increases so that the
chemical potential becomes substantially larger than the mass of
the $s$-quark. Here we consider only the $T=0$ case. The possible
coexistence of the chiral and diquark condensates is neglected on
the basis of the Anderson theorem \cite{9}. The expression for the
thermodynamic potential reads \cite{2,9}: $$
\Omega(T=0,\mu;\,\Delta)=\frac{\Delta^2}{4g}-\frac{2}{\pi^2}\int
dqq^2\left\{\sqrt{(E_q - \mu)^2+ \Delta^2}+\right.$$
\begin{equation}
\left.+\sqrt{(E_q +\mu)^2 +\Delta^2}+|\mu-E_q |+ E_q\right\},
\label{1}
\end{equation}
where $E_q=\sqrt{q^2+m^2}$.

The four-fermion interaction constant $g$ has a dimension
$m^{-2}$. From (\ref{1}) one obtains the following gap equation:
\begin{equation}
\Delta=\frac{4g}{\pi^2}\int dqq^2 \left(\frac{\Delta}{\mathcal{E}_q}+
\frac{\Delta}{\overline{\mathcal{E}_q}}\right),
\label{2}
\end{equation}
where $\mathcal{E}_q = \sqrt{(E_q - \mu)^2 +\Delta^2}$,
$\overline{\mathcal{E}_q} = \sqrt{(E_q +\mu)^2+\Delta^2}$.

An important remark concerning the structure of Eqs. (\ref{1}),
(\ref{2}) is due here. We have tacitly assumed that the
four-fermion interaction between quarks is point-like. In general
case instead of (2) one should write
\begin{equation}
\Delta_p = \frac{4g}{\pi^2}\int dqq^2
V_{pq}\left(\frac{\Delta_q}{\mathcal{E}_q}+
\frac{\Delta_q}{\overline{\mathcal{E}_q}}\right).
\label{3}
\end{equation}

However, the crude approximation (\ref{1}), (\ref{2}) is suffice
to obtain the general structure of the Klein--Gordon equation.
Next, making use of the standard Bogolubov functional, we
introduce the wave functions of the quark--quark and
antiquark--antiquark pairs $$\varphi_p =
\frac{\Delta}{\mathcal{E}_p},\quad \chi_p =
\frac{\Delta}{\overline{\mathcal{E}_p}}.$$

Then with a little juggling of (\ref{2}), (\ref{3}) we obtain the
following set of coupled equations for $\varphi_p$ and $\chi_p$:

\begin{equation}(\sqrt{p^2+ m^2}-{\mu})\varphi_p=\frac{4g}{\pi^2}(1-2n_p)\int
dqq^2(\varphi_q+\chi_q),\label{4} \end{equation}
\begin{equation}(\sqrt{p^2+m^2}+\mu)\chi_p=\frac{4g}{\pi^2}(1-2\bar{n}_p)\int
dqq^2(\chi_q+\varphi_q),\label{5} \end{equation}where
\begin{equation}1-2n_p= \frac{E_p-\mu}{\mathcal{E}_p},\quad
1-2\bar{n}_p=\frac{E_p+\mu}{\overline{\mathcal{E}_p}}.\label{6}
\end{equation}

These two equations may be recasted into a single Klein--Gordon
equation following the standard procedure \cite{10}. Let us define
$\psi_p=\varphi_p+\chi_p,$ and consider the dilute limit $n_p\ll1,
\bar{n}_p\ll1.$

Then one gets
\begin{equation} (p^2+m^2-\mu^2)\psi_p=\frac{8 g}{\pi^2}\sqrt{p^2+m^2}\int dqq^2 \psi_q. \label{7}
\end{equation}

We remind that non-locality in (\ref{7}) is of a symbolic
character as soon as we use the point-like four-fermion
interaction.

From (\ref{4}), (\ref{5}), and (\ref{7}) we see that in the dilute
strong coupling limit the chemical potential $\mu$ plays the role
of the eigenvalue of the Klein--Gordon equation. The point
$\mu=m-\mathcal{E}_B$ corresponds to negative nonrelativistic
chemical potential, typical for the ``molecular'' limit of the
BCS--BEC crossover. The phase diagram in the $(n_p/\bar{n}_p,\mu)$
plane has two symmetric branches corresponding to quarks and
antiquarks. Note also that the system described by Eqs. (\ref{4}),
(\ref{5}) and (\ref{7}) possesses the ``exiton-like'' instability.

The author is grateful for discussions and remarks from
N.O.~Agasian, T.D.~Lee, E.V.~Shuryak, D.T.~Son, M.A.~Stephanov,
and A.M.~Tsvelik. Special thanks to T.Yu.~Matveeva for the help in
preparing the article. Financial support from BNL and Grant
Ssc--1774--2003 is gratefully acknowledged. We thank INT (Seattle)
for its hospitality and the Department of Energy for the support
during the Workshop INT--04--1.

It is a pleasure and a honor to submit this paper to Yurii Simonov
Festschrift.


\end{document}